\documentclass[a4paper]{IEEEtran}
 
\usepackage{amsmath, amsfonts, amssymb}
\usepackage{bbm}
\usepackage{bbold}
\usepackage{graphicx,dblfloatfix}
\usepackage[space]{cite}
\usepackage{hyperref}
\usepackage{color}
 
\newtheorem{theorem}{Theorem}

\newtheorem{proposition}[theorem]{Proposition}
\newtheorem{definition}{Definition}

\newcommand{\mbf}[1]{\mathbf{#1}}
\newcommand{\set}[1]{\mathcal{#1}}

\newcommand{\bfZ}{\mathbb{Z}}

\renewcommand{\Pr}{\mathbb{P}}

\newcommand{\M}{\set{M}}
\renewcommand{\d}{\mbf{d}}
\newcommand{\Pe}{{\mathsf{P}_\text{e}}}

\newcommand{\mw}[1]{{\color{black}{#1}}}

\allowdisplaybreaks[4] 
\graphicspath{{.}{./Figures/}} 

\IEEEoverridecommandlockouts

\begin{document}
\title{Joint Cache-Channel Coding over Erasure Broadcast Channels}
\author{Roy Timo and Mich\`{e}le Wigger
\thanks{
R.~Timo is with TU M\"{u}nchen (roy.timo@tum.de), and M.~Wigger is with Telecom ParisTech (michele.wigger@telecom-paristech.fr). This work was supported by the Alexander von Humboldt Foundation.}
}

\maketitle

\begin{abstract}
We consider a cache-aided communications system in which a transmitter communicates with many receivers over an erasure broadcast channel. The system serves as a basic model for communicating on-demand content during periods of high network congestion, where some content can be pre-placed in local caches near the receivers. We formulate the cache-aided communications problem as a joint cache-channel coding problem, and characterise some information-theoretic tradeoffs between reliable communications rates and cache sizes. We show that if the receivers experience different channel qualities, then using unequal cache sizes and joint cache-channel coding improves system efficiency.
\end{abstract}


\section{Introduction} 

Consider a network with one transmitter and many receivers. Imagine that the transmitter has a library of \emph{messages} (or, \emph{data files}), and suppose that each receiver will request and download a message during a period of high network congestion. In such settings, it is advantageous to move traffic away from the congested period using \emph{caching}. The basic idea of caching is that the transmitter sends and stores ``parts'' of the library in local \emph{cache memories} near the receivers beforehand, during periods with low network traffic. The caches provide this data directly to the receivers, so that less data needs to be sent during the congested period. 

The above problem is relevant to video-streaming services, where content providers pre-place data in clients' caches (or, on servers near the clients), with the goal of improving latency and rate performance in high demand periods. The content provider typically does not know in advance which specific movies the clients will request, and thus the cached data cannot depend on the clients' specific demands.

Let us call the pre-placement of data in caches the \emph{caching phase}, and the remaining communications phase the \emph{delivery phase}. Cache memories are typically much smaller than the library, and the caching phase occurs before the receiversÕ demands are known. A key engineering challenge is, therefore, to carefully choose and cache only that data which is most useful during the delivery phase. That is, one should cache data that minimises the rate needed to complete the delivery-phase downloads for any feasible receiver demands. 

Cache-aided communications systems have received significant attention in the information-theoretic literature in recent years, and those works most closely related this paper are\cite{maddahali_niesen_2014-1,maddahali_niesen_2014-2,niesen_maddahali_2014,wang_xian_liu_2015,pedarsani_maddahali_niesen_2015,karamchandani_niesen_maddah_ali_diggavi_2014,hachem_karamchandani_diggavi_2014,ghasemi_ramamoorthy,wang_lim_gastpar_2015, ji_tulino_llorca_caire_2015, huang_wang_ding_yang_zhang_2015}.  With the exception of~\cite{huang_wang_ding_yang_zhang_2015}, these works assume that the delivery phase takes place over a single rate-limited multicast noiseless channel (a bit-pipe) that connects the transmitter to every receiver. In practice, however, the communications medium is sometimes better modelled by a noisy broadcast channel (BC). This scenario is considered in \cite{huang_wang_ding_yang_zhang_2015}, where the BC  is essentially a set of parallel links with different qualities to the various receivers, which models a wireless fading BC. 

This paper takes a similar approach to that of~\cite{huang_wang_ding_yang_zhang_2015}, and we assume that the delivery phase takes place over a memoryless erasure BC. However, in contrast to~\cite{huang_wang_ding_yang_zhang_2015}, we assume that the caching phase takes place over error-free pipes. The motivation for this simplified assumption is that the caching phase typically occurs during periods of low network-congestion, where network resources are not a limiting factor. 

Our main contribution in this paper is a joint cache-channel coding scheme for the described setup for general demands, and a characterization of the \emph{capacity-memory} region when the receivers wish to learn the same message. Our results show that when the receivers experience different erasure probabilities (different channel qualities), then
\begin{itemize}
\item it is beneficial to employ unequal cache sizes at the receivers (larger cache memories at weaker receivers, and smaller cache memories at strong receivers); and
\item joint cache-channel coding techniques can provide significant gains over separated cache and channel coding.
\end{itemize}

Allocating larger cache memories to the weaker receivers is quite natural because one then needs to communicate less data over noisier channels (see also \cite{huang_wang_ding_yang_zhang_2015}). Interestingly,  there is an additional benefit to asymmetric caches that arises when joint cache-channel coding is used during the delivery phase. The basic idea is as follows: Consider a degraded BC communications scenario (such as the erasure BC) with separate cache and channel coding. Here a stronger receiver can decode all the data that is sent to a weaker receiver during the delivery phase. In fact, the strong receiver could decode even more data, but it is limited by the weaker receiver. Now suppose that part of the message intended for the stronger receiver is stored within the weaker receiver's cache: one can freely piggyback this part of the stronger receiver's message on the message intended for the weaker receiver. The weaker receiver is not penalised because it knows what data is being piggybacked on its desired message, and its channel decoder can still resolve its desired message. While, simultaneously, the stronger receiver has decoded something about its desired message and therefore we have improved efficiently. Thus, thanks to the weaker receiver's cache and a simple joint cache-channel coding scheme, we can send additional data to stronger receivers without any extra cost, i.e., extra rate-constraints. This additional benefit of asymmetric cache memories was not observed in \cite{huang_wang_ding_yang_zhang_2015}, because a separate source-channel coding scheme was used for the delivery phase.


\section{Problem Definition}


\subsection{Message library and feasible receiver demands}

We have a transmitter, $K$ receivers and a library with $D$ messages $W_1,\ldots, W_D$. The $d$-th message in the library $W_d$ is independent of all other messages and uniform on\footnote{To simplify notation and help elucidate our main ideas, we assume throughout the paper that $2^{nR_d}$ is an integer.}
\begin{equation*}
\big\{0,1,\ldots, 2^{n R_d} - 1 \big\},
\end{equation*}
where $R_d \geq 0$ is its rate and $n$ is the transmission blocklength. We represent a particular combination of receivers' demands by a tuple $\d = (d_1,\ldots, d_K) \in \{1,\ldots, D\}^K.$ That is,~$\d$ represents the situation where receiver~$1$ demands (i.e., requests and downloads) message $W_{d_1}$, receiver~$2$ demands $W_{d_2}$, and so on. Let 
\begin{equation*}
\set{D}\subseteq \{1,\ldots, D\}^K.
\end{equation*}
denote the \emph{feasible set} of all possible receiver demands. The feasible set $\set{D}$ is known to the transmitter and receivers during the caching and delivery phases, but the specific demand tuple $\d$ chosen from $\set{D}$ is only revealed for the delivery phase. 


\subsection{Caching phase} 

For each receiver $k \in \{1,\ldots,K\}$,  the size of its cache is described by a nonnegative integer $\M_k$, see \eqref{eq:cachesize} below. The transmitter sends 
\begin{equation*}
\bfZ_k := g_k(W_1, \ldots, W_D),
\end{equation*}
to receiver~$k$'s cache, where 
$g_k : \prod_{d = 1}^D \{0,1,\ldots, 2^{nR_d} - 1\} 
\to \set{Z}_k$.
such that 
\begin{equation} \label{eq:cachesize}
\log | \set{Z}_k| \leq 2^{n \M_k}. 
\end{equation}

The caching phase occurs during a low congestion period, and we assume that $\bfZ_k$ is reliably conveyed to receiver~$k$'s cache (for each $k \in \{1,\ldots,K\}$). 


\subsection{Erasure Broadcast Channel Model} 

The delivery phase occurs during a high congestion period, which we model by an \emph{erasure BC} with input alphabet $\set{X} := \{0,1\}^{F}$. Here $F \geq 0$ is a fixed positive integer, and each $x \in \set{X}$ is an $F$-bit packet. Due to congestion, some packets may be lost when, for example,  router buffers overload. We denote the event of a lost packet with the \emph{erasure symbol} $\Delta$, and the BC's output alphabet by $\set{Y} := \set{X} \cup \{\Delta\}$ (the same alphabet is used for all receivers). Fix  
\begin{equation*}
1 \geq \delta_1 \geq \delta_2 \geq \cdots \geq \delta_K \geq 0.
\end{equation*}
Let $Q(y_1,\ldots,y_k|x)\! :=\! \Pr[(Y_1,\ldots,Y_K)\! =\!(y_1,\ldots,y_K)|X\!=\!x]$
be any probability law for the memoryless BC with marginals
\begin{equation*}
\Pr[Y_k = y_k |X = x] = 
\left\{
\begin{array}{cl}
1 - \delta_k & \text{ if } y_k = x\\
\delta_k & \text{ if } y_k = \Delta\\
0 & \text{ otherwise}
\end{array}
\right.,
\quad \forall\ k.
\end{equation*}
For our purpose only these marginal probabilities are relevant.

We discuss a caching system in the next section that is built on separate cache and channel codes, and, for this reason, it is useful to recall the degraded message set capacity region for $Q$. A channel-coding rate tuple $(R_{\{1,\ldots,K\}},$ $R_{\{2,\ldots,K\}},\ldots,R_{\{K\}})$ is said to be achievable on $Q$ if the following holds: For any $\epsilon > 0$ there exists an encoder and $K$-decoders such that, for all $k$,  the transmitter can send $(R_{\{k,\ldots,K\}} - \epsilon)$ information bits per channel use to every receiver in the set $\{k,k+1,\ldots,$ $K\}$ with an average probability of error less than $\epsilon$. The set of all achievable rates --- the \emph{degraded message set capacity region} $\set{C}^\dag$ --- is given by the next proposition. The proposition can be distilled from~\cite{Urbanke-1999-A}, and we omit these details.  

\begin{proposition}\label{Prop:CapacityRegion} 
\begin{multline*}
\set{C}^\dag = \Bigg\{
\big(R_{\{1,\ldots,K\}},R_{\{2,\ldots,K\}},\ldots,R_{\{K\}}\big) :
\\
\left.
\sum_{k=1}^K 
\frac{R_{\{k,\ldots,K\}}}{F(1-\delta_k)} 
\leq 1,\ R_{\{k,\ldots,K\}} \geq 0,
\ \ \forall\ k
\right\}.
\end{multline*}
\end{proposition}


\subsection{Delivery phase}

For each feasible demand $\d \in \set{D}$, let
\begin{equation*}
f_\d : \prod_{d' = 1}^D 
\{0,1,\ldots, 2^{nR_{d'}}-1\}
\to 
\set{X}^n
\end{equation*}
denote the corresponding encoder at the transmitter. Given $\d \in \set{D}$ and the library $(W_1,\ldots,W_D)$, the transmitter sends 
\begin{equation}\label{eq:enc}
X^n := f_\d(W_1,\ldots,W_D),
\end{equation}
where $X^n = (X_1,\ldots,X_n)$. Receiver $k$ observes $Y_k^n = (Y_{k,1},$ $\ldots,Y_{k,n})$ according to the memoryless law $Q$. Let 
\begin{equation}\label{eq:decod}
\varphi_{k,\d} : \set{Y}^n \times \set{Z}_k \to \{0,1,\ldots, 2^{nR_{d_k}}-1\}
\end{equation}
denote the decoder at receiver $k$. Given demands $\d \in \set{D}$, cache content $\bfZ_k$ and channel outputs $Y_k^n$, receiver $k$ outputs 
\begin{equation*}
\hat{W}_k := \varphi_{k,\d}(Y_k^n, \bfZ_k)
\end{equation*}
as its reconstruction of the $d_k$-th message $W_{d_k}$. 

\subsection{Achievable rate-memory tuples}
Let
\begin{equation*}
\Pe := 
\Pr\bigg[\ \bigcup_{\d \in \set{D}} 
\bigcup_{k = 1}^K
\big\{ \hat{W}_k \neq W_{d_k} \big\}\
\bigg]
\end{equation*}
denote the probability of error at any receiver for any feasible demand.  We call the collection of all encoders and decoders, 
\begin{equation*}
\big\{g_1,g_2,\ldots,g_K\big\}
\text{ and }  
\big\{f_\d,\varphi_{1,\d},\varphi_{2,\d},\ldots,\varphi_{K,\d}\big\}_{\d \in \set{D}},
\end{equation*} 
an $(n,R_1,\ldots,R_D,\M_1,\ldots,\M_K)$-code. 

We say that a rate-memory tuple $(R_1,\ldots,R_D,\M_1,\ldots,$ $\M_K)$ is \emph{achievable} if for any $\epsilon >  0$ there exists a sufficiently large blocklength $n$ and an $(n,R_1,\ldots,R_D,\M_1,$ $\ldots,\M_K)$-code with $\Pe \leq \epsilon$. 

\begin{definition}
We define the \emph{capacity-memory} region $\set{C}$ to be the closure of the set of all achievable rate-memory tuples.
\end{definition}

The main problem of interest in this paper is to determine the capacity-memory region $\set{C}$ for a given erasure BC $Q$. 


\section{Motivating Examples}

We now demonstrate the potential of unequal cache memories and joint cache-channel coding with three examples. Fix $K=2$; $\set{D} = \{1,\ldots,D\}^2$; $R_d = R$ for all $d$; and 
\begin{equation}\label{eq:channels}
\delta_1=4/5
\quad
\text{and}
\quad
\delta_2=1/5.
\end{equation}


\subsection{Coded caching with symmetric caches}\label{Sec:codedcachingsymmetric}

Suppose that $\M_1=\M_2 = \M$, and
\begin{equation}\label{eq:condalpha}
\alpha := \M/R \in [0, D/2].
\end{equation}
Split each message $W_d$ in the library into three sub-messages, 
\begin{equation*}
W_d = \big(W_{d}^{\textnormal{(c1)}},W_{d}^{\textnormal{(c2)}},W_{d}^{\textnormal{(u)}}\big),
\end{equation*}
of rates ${\M}/{D}$, ${\M}/{D}$, and $R- {2\M}/{D}$.

\textit{Caching phase:} 
Store the sub-messages 
\begin{equation*}
(W_{1}^{\textnormal{(c1)}},\ldots,W_{D}^{\textnormal{(c1)}})
\quad
\text{and}
\quad
(W_{1}^{\textnormal{(c2)}},\ldots,W_{D}^{\textnormal{(c2)}})
\end{equation*}
in the caches of receiver~$1$ and $2$'s respectively.

\textit{Delivery phase:} 
The transmitter sends 
\begin{equation}\label{Eqn:XOR-Mess1}
W_{d_1}^{\textnormal{(c2)}} \oplus W_{d_2}^{\textnormal{(c1)}},
\end{equation} 
as a common message to both receivers, where the addition is modulo $2^{n (\M/D)}$. It then sends $W_{d_1}^{\textnormal{(u)}}$ as a private message to receiver~$1$ and  $W_{d_2}^{\textnormal{(u)}}$ as a private message to receiver~$2$. Notice that receiver $1$ can recover $W_{d_1}$ from the common message and $W_{d_1}^{\textnormal{(u)}}$, while receiver~$2$ can recover $W_{d_2}$ from the common message and $W_{d_2}^{\textnormal{(u)}}$. We use a good channel code to communicate these messages over the BC.

\textit{{Achievable rate-memory tuples:}} Proposition~\ref{Prop:CapacityRegion} asserts that the common message~\eqref{Eqn:XOR-Mess1} and $W_{d_1}^{\textnormal{(u)}}$ can be decoded by both receivers and  $W_{d_2}^{\textnormal{(u)}}$ can be decoded by receiver~$2$ whenever
\begin{equation}\label{eq:trad0}
\frac{R -  \frac{\M}{D}}{F(1-\delta_1)} + \frac{R - \frac{2\M}{D}}{F(1-\delta_2)} \leq 1.
\end{equation} 
On substituting~\eqref{eq:channels}, the inequality~\eqref{eq:trad0} simplifies to
\begin{equation}\label{eq:trade}
R \leq \frac{4}{5} F(1-\delta_1) + \frac{6}{5} \frac{\M}{D}.
\end{equation}
All rate-memory tuples $(R,\ldots,R,\M,\ldots,\M)$, with $R$ and $M$ satisfying~\eqref{eq:condalpha} and~\eqref{eq:trade}, are achievable.

 
\subsection{Separate cache-channel coding and asymmetric caches}\label{Sec:codedcachingasymmetric}

Now suppose that we have asymmetric caches 
 $\M_1 = 2 \M$ and $\M_2 = 0$ 
for some $\M$ satisfying~\eqref{eq:condalpha}. The total cache memory available at both receivers remains unchanged, only now the memory at receiver~$2$ has been reallocated to receiver~$1$. 

Split each message $W_d$ into two sub-messages, 
\begin{equation}\label{Eqn:CodedCaheSubmessages}
W_d = (W_{d}^{\textnormal{(c1)}},W_{d}^{\textnormal{(u)}})
\end{equation}
with rates $2 \M/D$ and $R-(2\M/D)$ respectively. 

\textit{Caching phase:} 
Store $(W_{1}^{\textnormal{(c1)}},\ldots,W_{D}^{\textnormal{(c1)}})$ in receiver~$1$'s cache. 

\textit{Delivery phase:} We use a good channel code for Proposition~\ref{Prop:CapacityRegion} to reliably communicate the above sub-messages. The transmitter sends $W_{d_1}^{\textnormal{(u)}}$ as a common message to both receivers (although it is only used by receiver~$1$), and it sends $W_{d_2}^{\textnormal{(c1)}}$ and  $W_{d_2}^{\textnormal{(u)}}$ as a private message to receiver~$2$. 


%
%

\textit{Achievable rate-memory tuples:}
Proposition~\ref{Prop:CapacityRegion} asserts that reliable communication is possible if 
\begin{equation}\label{eq:trad2}
\frac{R -  \frac{2\M}{D}}{F(1-\delta_1)} + \frac{R}{F(1-\delta_2)} \leq 1.
\end{equation} 
On substituting~\eqref{eq:channels}, the inequality~\eqref{eq:trad2} simplifies to
\begin{equation}\label{eq:trade3}
R \leq \frac{4}{5} F(1-\delta_1) + \frac{8}{5} \frac{\M}{D}.
\end{equation} 
All rate-memory tuples $(R,\ldots,R,\M,\ldots,\M)$, with $R$ and $\M$ satisfying~\eqref{eq:condalpha} and~\eqref{eq:trade3}  are achievable. 


\subsection{Joint cache-channel coding and asymmetric caches}\label{Sec:ExampleJSCC}

As in Section~\ref{Sec:codedcachingasymmetric}: Let $\M_2 = 0 \text{ and } \M_1 = 2 \M$, for some $\M$ satisfying~\eqref{eq:condalpha}, and split each message $W_d$ into two sub-messages~\eqref{Eqn:CodedCaheSubmessages} with rates $2 \M/D$ and $R-(2\M/D)$ respectively. 

\textit{Caching phase:} 
Store $(W_{1}^{\textnormal{(c1)}},\ldots,W_{D}^{\textnormal{(c1)}})$ at receiver~$1$. 

\textit{Delivery phase:} Transmission takes place in two phases using \mw{timesharing}. First phase of length $\beta_1 n$, for some $\beta_1\in[0,1]$: The transmitter sends 
\begin{equation*}
(W_{d_1}^{\textnormal{(u)}} ,  W_{d_2}^{\textnormal{(c1)}})
\end{equation*} 
as a common message to both receivers using a joint cache-channel code. Second phase of length $(1-\beta_1)n$" The transmitter sends $W_{d_2}^{\textnormal{(u)}}$ to receiver~2 using a point-to-point channel code. Receiver~1 tries to decode $W_{d_1}^{\textnormal{(u)}}$ and receiver~2 tries to decode $(W_{d_1}^{\textnormal{(u)}},W_{d_2}^{\textnormal{(c1)}},W_{d_2}^{\textnormal{(u)}})$. A key observation here is that $W_{d_2}^{\textnormal{(c1)}}$ is stored in receiver~$1$'s cache. \mw{As we see in a moment, for $\alpha\in\{0, \frac{3D}{8}\}$, this allows to freely piggyback} 
receiver~$2$'s message $W_{d_2}^{\textnormal{(c1)}}$ 
on receiver~$1$'s message $W_{d_1}^{\textnormal{(u)}}$ without compromising the rate to receiver~$1$. 


\textit{Achievable rate-memory tuples:} 
\mw{By Tuncel's  seminal \emph{broadcasting with side-information} result \cite{Tuncel-Apr-2006-A},  communication in phase~$1$ (to both receivers) is very likely to be successful  if the following two conditions hold:
\begin{subequations}\label{eq:cc}
\begin{IEEEeqnarray}{rCl}
R - \frac{2\M}{D} & \leq & F(1-\delta_1) \beta_1\\
R & \leq & F( 1-\delta_2) \beta_1;
\end{IEEEeqnarray}
communication in  phase $2$  is very likely to be successful if
\begin{equation}
R - \frac{2\M}{D} \leq F(1-\delta_2) (1-\beta_1).
\end{equation}
\end{subequations}
Inequalities~\eqref{eq:cc} prove achievability of all rate-memory tuples $(R,\ldots,R,\M,\ldots,\M)$, with $R$ and $\M$ satisfying~\eqref{eq:condalpha} and
\begin{equation}\label{eq:trade5}
 R \leq \begin{cases} \frac{4}{5} F(1-\delta_1) +2 \frac{\M}{D}, & \textnormal{ if } \frac{\M}{R} \in \big[ 0, \frac{3D}{8} \big] \\
2 F (1- \delta_1) + \frac{\M}{D} & \textnormal{ if } \frac{\M}{R} \in \big(  \frac{3D}{8}, \frac{D}{2} \big].  \end{cases}
\end{equation}
}

\subsection{Discussion}
Comparing the rate-memory tradeoffs in~\eqref{eq:trade}, \eqref{eq:trade3} and \eqref{eq:trade5}, we see that it is advantageous to use unequal cache sizes and joint cache-channel coding. In particular, allowing larger caches at the weaker receivers (with higher packet erasure probabilities) both reduces the delivery-phase rates to the weaker receivers and increases rates to the stronger receivers.


\section{A Joint Cache-Channel Code for\\ Arbitrary Demands}

We now describe a joint cache-channel code that can be applied for any set of feasible demands $\set{D}$, but we restrict attention to equal message rates 
\begin{equation*} 
R_d = R, \qquad d\in\{1,\ldots, D\}.
\end{equation*}  We first treat the case where the $K_0$ weakest receivers (receivers~$1$ to $K_0$) have equal cache sizes and the remaining receivers do not have caches:
\begin{equation}\label{eq:assumptions_general}
\M_k = 
\left\{
\begin{array}{cc}
\M & \text{ if } k \leq K_0\\
0 & \text{ if } k > K_0
\end{array}
\right. .
\end{equation}
We explain later in Section~\ref{sec:extension} how the scheme can be generalised to setups with unequal cache sizes.


\subsection{Scheme for cache sizes satisfying \eqref{eq:assumptions_general}}

\textit{Preliminaries:}
Choose a positive integer $t < K_0$, and let 
\begin{equation*}
\tau := {{K_0}\choose{t}}.
\end{equation*}

Split each message $W_d$ into $(\tau+1)$-sub-messages, 
\begin{equation*}
W_d = \big(W_d^{(1)}, \ldots, W_d^{(\tau+1)}\big),
\end{equation*}
where
\begin{equation*}
W_d^{(i)} \in \big\{0,1,\ldots, 2^{n R(i)} - 1 \big\}
\end{equation*}
and 
{\arraycolsep=1.4pt\def\arraystretch{2.2}
\begin{equation*}
R^{(i)} := 
\left\{
\begin{array}{ll}
\dfrac{\M}{D {{K_0-1}\choose{t-1}}}, 
& \text{ for } i = 1,2,\ldots,\tau\\
R- \dfrac{\M K_0}{D t}, 
& \text{ for } i = \tau + 1.
\end{array}
\right.
\end{equation*}
} 

\textit{Caching Phase:} 
Consider the $K_0$ weakest receivers. Let 
\begin{equation*}
\set{R}_{1}, \set{R}_2, \ldots, \set{R}_\tau
\end{equation*}
denote the $\tau$ different subsets of $\{1,\ldots,K_0\}$ with size $t$. For each $i = 1,2,\ldots,\tau$, take the tuple
\begin{equation*}
(W_1^{(i)}, W_2^{(i)}, \ldots, W_D^{(i)})
\end{equation*}
and store it in the cache of each and every receiver in $\set{R}_i$. Here we have stored $D{K_0-1 \choose t-1}$ sub-messages in receiver~$k$'s cache (for each $k \in \{1,2,\ldots,K_0\}$) with a total memory requirement 
\begin{equation*}
\left(2^{n \frac{\M}{D} {K_0-1 \choose{t-1}}^{-1}} \right)^{D{K_0-1 \choose t-1}}
= 2^{n\M}.
\end{equation*}

\textit{Delivery phase:} The demand tuple $\d \in \set{D}$ is given, and we are required to communicate message $W_{d_1}$ to receiver~$1$, $W_{d_2}$ to receiver~$2$, and so on. 

We consider sets of $(t+1)$-receivers in $\{1,\ldots, K_0\}$. Within these sets, each subset of $t$ receivers shares a sub-message that is demanded (but unknown) by the remaining $(t+1)$-th receiver. For each set of $(t+1)$-receivers, we  form the ``XOR" of the $(t+1)$ sub-messages having the two above mentioned properties, that is, being known at $t$ of the receivers and demanded by the remaining $(t+1)$-th receiver. For example, for the subset of  receivers~$\{1, \ldots, t+1\}$, we form the XOR message
\begin{equation*}
\bigoplus_{k=1}^{t+1} W_{d_k}^{(i_k)},  
\end{equation*}
where the addition is modulo $2^{\M/(D{K_0-1\choose t-1})}$ (or, equivalently, a bitwise XOR operation); and for each $k\in\{1,\ldots, t+1\}$,  $i_k$ is such that
\begin{equation}\label{eq:cod_msg}
\set{R}_{i_k} \triangleq \{1,\ldots, k-1, k+1, \ldots, t+1\}. 
\end{equation} 
Notice that~\eqref{eq:cod_msg}  implies that $W_{d_k}^{(i_k)}$  is stored in the caches of receivers~$1,\ldots, k-1, k+1, \ldots, t$, but not at receiver~$k$.

We use a time-sharing scheme to send the XOR messages as well as all other messages to be transmitted. The time-sharing comprises $K$ phases. Each phase~$k \in \{1,\ldots,K\}$ is constructed so that it can be decoded by Receivers~$k, k+1, \ldots, K$. Phase~$k$ is of length  $\beta_k n$, where\begin{equation}\label{Eqn:Beta}
\sum_{k=1}^K   \beta_k=1, \quad 0 \leq \beta_k \leq 1.
\end{equation}
In phase~$k\in\{1,\ldots, K_0\}$, we send
\begin{itemize}
\item  the XOR messages that are demanded by receiver~$k$ but not by receivers~$1$ to $k-1$;
\item the uncached message $W_{d_k}^{(\tau+1)}$ demanded by Receiver~$k$;
\item the first $n C_{k,\tilde{k}}$ bits of sub-messages $W_{d_{\tilde{k}}}^{(i)}$, for every $\tilde{k} \in\{K_0+1,\ldots, K\}$ and every $i\in\{1,\ldots,\tau\}$ such that $k\in\set{R}_i$. These messages are all known to receiver~$k$ and therefore do not limit the decoding at receiver~$k$. The rates $\{C_{k, \tilde{k}}\}$ are parameters of a scheme. As we shall see, when they are chosen sufficiently small, but positive, and $\delta_{k+1}< \delta_k$, then sending these bits does not limit the decoding at receivers~$k+1, k+2, \ldots, K$. In fact, similarly to our motivating example, in this case, the transmitted bits of sub-messages $W_{d_{\tilde{k}}}^{(i)}$ can be freely piggybacked on the other messages transmitted in this phase~$k$.
\end{itemize}
In phase~$k\in\{K_0+1,\ldots, K\}$ we send: 
\begin{itemize}
\item the sub-messages of $W_{d_{k}}$ that have not been sent in any previous phase.
\end{itemize}
\textit{Achievable rate-memory tuples:}  
\begin{proposition}A rate-memory tuple  $(R, \ldots, R, \M_1=\M,$ $\ldots, \M_{K_0}=\M, 0, \ldots, 0)$ is achievable if  for some
 \begin{itemize}
 \item  positive integer $t$;
 \item  nonnegative $K$-tuple $(\beta_1,\ldots, \beta_K)$ satisfying~\eqref{Eqn:Beta};  and 
 \item  nonnegative real numbers $\{C_{k, \tilde{k}}\}$ with $k\in\{1,\ldots, K_0\}$ and $\tilde{k}\in\{K_0+1,\ldots, K\}$;
 \end{itemize}
the  following $(K+K_0)$-conditions in~\eqref{eq:conditions} hold.
\begin{enumerate}
\item
 For each $k\in\{1,\ldots, K_0-t-1\}$, we have
  \begin{subequations}\label{eq:conditions}
 \begin{IEEEeqnarray}{rCl}\label{eq:cond1}
 R 
 & \leq & F (1-\delta_k)+ \frac{\M}{D{K_0-1\choose t-1}}\left( {K_0 \choose t} - {K_0-k \choose t}\right) \nonumber \\ 
  \end{IEEEeqnarray}
 and 
\begin{align}
\notag
R\ + \sum_{\tilde{k}=K_0+1}^K & C_{k,\tilde{k}} \leq F (1-\delta_{k+1})
\\
\label{eq:cond2}
&
+ \frac{\M}{D{K_0-1\choose t-1}}\left( {K_0 \choose t} - {K_0-k \choose t}\right).
\end{align}

\item For each $k\in\{ K_0-t, \ldots, K_0\}$, we have
  \begin{IEEEeqnarray}{rCl}\label{eq:cond3}
  R 
 & \leq & F (1-\delta_{k})+ \frac{\M K_0}{Dt}
   \end{IEEEeqnarray}
 and 
  \begin{IEEEeqnarray}{rCl}\label{eq:cond4}
R + \sum_{\tilde{k}=K_0+1}^K C_{k,\tilde{k}}
 & \leq & F (1-\delta_{k+1})+ \frac{\M K_0}{Dt}.\IEEEeqnarraynumspace
 \end{IEEEeqnarray}
 
\item Finally, for each $k\in\{K_0+1,\ldots, K\}$, we have 
 \begin{IEEEeqnarray}{rCl}\label{eq:cond5}
R - \sum_{k'=1}^{K_0} C_{{k}',k}  \leq F(1-\delta_k).\IEEEeqnarraynumspace
 \end{IEEEeqnarray}
 \end{subequations}
 \end{enumerate}
 \end{proposition}
 
\begin{IEEEproof}[Proof outline]
For each $k \in \{1,\ldots, K_0-t\}$, Condition~\eqref{eq:cond1} ensures that receiver $k$ can reliably decode the sub-messages sent during phase~$k$, and Condition~\eqref{eq:cond2} ensures that all of the stronger receivers in $\{k+1, \ldots, K\}$ can also reliably decode these sub-messages. Similarly, Condition~\eqref{eq:cond3} ensures that each receiver $k \in \{K_0-t, \ldots, K_0\}$ can reliably decode the sub-messages sent in phase~$k$, and Condition~\eqref{eq:cond4} ensures that all of the stronger receivers in~$\{k+1, \ldots, K\}$ can also reliably decode these sub-messages. Finally, Condition~\eqref{eq:cond5} ensures that each receiver~$k\in\{K_0+1, \ldots, K\}$ can decode the sub-messages sent in phase~$k$. 
\end{IEEEproof}

 \textit{Discussion:} The parameters $\{C_{k, \tilde{k}}\}$ describe the gain  that our scheme achieves over separate cache-channel coding schemes. If some of these rates are strictly larger than 0, then our scheme strictly outperforms separate cache-channel coding. It is possible to choose them strictly positive whenever the erasure probabilities $\delta_1, \ldots, \delta_{K_0}$ are not all equal.
 
We took advantage of the fact that receiver~$k$ has already cached the additional $nC_{k,\tilde{k}}$ message bits that are sent in phase~$k$. Some of these bits are also available to the next-stronger receivers~$k+1, k+2, \ldots$. For simplicity, we ignored this fact in our analysis, and it is likely that further gains can still be made.

  
 \subsection{Scheme for unequal cache sizes}\label{sec:extension} Assume now that 
 \begin{equation}
 \M_1\geq \M_2 \geq \cdots \M_K \geq 0.
 \end{equation}
Our scheme in the previous subsection is easily extended to this more general setup using time-sharing. Specifically: Let $\beta_1, \ldots, \beta_K$ be real numbers  in the interval $[0,1]$ that sum up to 1. Over a fraction of time $\beta_i$, $i\in\{1,\ldots, K\}$, we use our scheme in the previous subsection assuming that only the first $K_{0}^{(i)}=K+1-i$ receivers have caches of equal cache size $\set{M}^{(i)}=\beta_i^{-1}(\set{M}_{K-i+1}- \set{M}_{K-i+2})$. (Set $\set{M}_{K+1}\triangleq 0$.) 

\section{Single Common Demand}\label{sec:common}
In this section we consider the optimistic case where all receivers demand the same message. This corresponds to
\begin{equation*}\label{eq:common}
\set{D}= \big\{ (d_1, \ldots, d_K) \in\{1,\ldots, D\}^{K}\colon d_1=d_2=\cdots=d_K\big\}.
\end{equation*} 
The rates $R_1, \ldots, R_D$ can be arbitrary, i.e., do not have to be equal as in the previous section.

\subsection{Result}
\begin{theorem}\label{thm:common} A rate-memory tuple  $(R_1, \ldots, R_D, \M_1,\ldots,$ $ \M_K)$ is achievable if and only if, 
\begin{IEEEeqnarray}{rCl}\label{eq:RdM}
R_d \leq \min_{k\in\{1,\ldots, K\}}\big( ( 1- \delta_k)F + \M_{k,d} \big), \ \  d\in\{1,\ldots, D\},\IEEEeqnarraynumspace
\end{IEEEeqnarray}
 for some nonnegative numbers $\{\M_{k,d}\}$  that satisfy
\begin{equation}
\sum_{d=1}^D \M_{k,d} \leq \M_k, \quad k\in\{1,\ldots, K\}.
\end{equation} 
\end{theorem} 
\begin{IEEEproof}
See the following two subsections.
\end{IEEEproof}
 We thus again wish to allocate small cache sizes to strong receivers  and large cache sizes to  weak receivers.

If we used separate cache-channel codes, Constraint~\eqref{eq:RdM} is replaced by
\begin{IEEEeqnarray}{rCl}
\max_{k\in\{1,\ldots, K\}} (R_d - \M_{k,d})\leq \min_{k\in\{1,\ldots, K\}}(  1- \delta_k)F, 
\end{IEEEeqnarray}
and the benefit of having unequal cache sizes $\{\M_{d}\}$ at the different receivers disappears.

\subsection{Proof of achievability}\label{sec:ach}
We propose the following scheme.

\textit{Caching phase:} 
Each receiver~$k$ stores in its cache the first $n\M_{k,d}$ bits of each Message~$W_d$, for $d\in\{1,\ldots, L\}$, where
\begin{equation}
\sum_{d=1}^N \M_{k,d} \leq \M_k,
\end{equation}
in order to satisfy the cache-memory constraint.

\textit{Delivery phase:} Assume $d_1=d_2=\ldots=d_K=d^\star$.
Use an i.i.d. Bernoulli-$1/2$ point-to-point code to send Message $W_{d^\star}$ to all receivers. Each receiver $k$ knows the first $n\M_{k,d^{\star}}$ bits of this message, and thus during its decoding it can restrict attention to the part of the codebook that corresponds to these bits. For receiver $k$ it is thus as if the transmitter had sent only its missing bits over the channel.  

Alternatively, a joint cache-channel code based on Tuncel's virtual binning technique~\cite{Tuncel-Apr-2006-A} can be used for the delivery phase. 

\textit{Achievable rate-memory tuples:}
By~\cite{Tuncel-Apr-2006-A}, whenever 
\begin{equation}\label{eq:Ckconstr}
R_d - \M_{k,d} \leq (1- \delta_k)F, \qquad \forall d\in\{1,\ldots, D\}, \; k\in\{1,\ldots, K\},
\end{equation}
the probability of error can be made arbitrarily small as $n\to \infty$. 

\subsection{Proof of Converse}\label{sec:con}
\allowdisplaybreaks[4]
Fix a block length $n$, and define
\begin{IEEEeqnarray}{rCl}
\M_{k,d} \triangleq\frac{1}{n} I(W_d; \bfZ_k), \quad k\in\{1,\ldots, K\}, \;d\in \{1,\ldots, D\}.\nonumber\\
\end{IEEEeqnarray}
For each $k\in\{1,\ldots, K\}$, 
\begin{IEEEeqnarray}{rCl}
\sum_{d=1}^N \M_{k,d} & = & \sum_{d=1}^D \frac{1}{n}I(W_d; \bfZ_k) \nonumber \\
&=&\frac{1}{n}\sum_{d=1}^D \big( H(W_d) -H(W_d | \bfZ_k) \big) \nonumber \\
 & = &\frac{1}{n}\big(  H(W_1, \ldots, W_D) -\sum_{d=1}^N H(W_d | \bfZ_k)\big)  \nonumber \\
 & \leq &\frac{1}{n} \big( H(W_1, \ldots, W_D) -H(W_1, \ldots, W_N | \bfZ_k)\big)  \nonumber \\
 & =&\frac{1}{n} I(W_1, \ldots, W_D; \bfZ_k) \nonumber \\
  &\leq &\frac{1}{n} H(\bfZ_k) \leq \M_k,
\end{IEEEeqnarray}
where the second and fourth equalities follow by the definition of mutual information; the third equality because the messages are independent; the first inequality because the sum of marginal entropies of a tuple of random variables, is at least as large as the joint entropy of this tuple. 

In the following, let $\epsilon_n$ denote any sequence that tends to 0 as $n\to \infty$. Fix  an achievable rate-memory tuple $(R_1, \ldots, R_N, \M_1, \ldots, \M_K)$. Also for an arbitrary large $n$, let $\{\bfZ_1,\ldots, \bfZ_K\}$, $\{f_{\mathbf{d}}\}$, and $\{\varphi_{k,\mathbf{d}}\}$ denote cache content,  encoding functions, and decoding functions achieving this rate-memory tuple. 
Fix now $d^\star\in\{1,\ldots, D\}$ and $k\in\{1,\ldots, K\}$, and 
let  $X^n=f_{\mathbf {d^\star}}(W_1,\ldots,W_D)$ and $Y^n$ denote inputs and outputs corresponding to demand $\mathbf{d}^\star\triangleq (d^\star, d^\star,\ldots, d^\star)$. We have
\begin{IEEEeqnarray}{rCl}
R_{d^\star} & \leq & \frac{1}{n} H(W_{d^\star}) \nonumber \\
& = & \frac{1}{n} I(W_{d^\star}; Y_k^n, \bfZ_k) + \frac{1}{n} H(W_{d^\star}|Y_k^n, \bfZ_k) \nonumber\\
& \leq & \frac{1}{n} I(W_{d^\star}; Y_k^n| \bfZ_k) + \frac{1}{n} I(W_{d^\star}; \bfZ_k) +   \epsilon_n\nonumber \\
& = & \frac{1}{n} \sum_{t=1}^n \big( H(Y_{k,t} | \bfZ_k, Y_{k}^{t-1}) - H(Y_{k,t} | W_{d^\star}, Y_{k}^{t-1}, \bfZ_k) \big)\nonumber \\
 & &+  \M_{k,{d^\star}} +\epsilon_n\nonumber \\ 
& \leq & \frac{1}{n} \sum_{t=1}^n \big( H(Y_{k,t}) - H(Y_{k,t} | X_{k,t})\big)+\M_{k,{d^\star}} + \epsilon_n\nonumber \\
& = & \frac{1}{n} \sum_{t=1}^n I(Y_{k,t}; X_{k,t})+ \M_{k,{d^\star}} +\epsilon_n\nonumber \\
& \leq & (1- \delta_k)F+ \M_{k,{d^\star}} +\epsilon_n,
\end{IEEEeqnarray} 
where the second inequality follows by Fano's inequality; the third inequality because conditioning cannot increase entropy and because of the Markov chain $(W_{d^\star}, Y_{k}^{t-1} \bfZ_k)\to X_{t} \to Y_{k,t}$; and the last inequality by the capacity of the erasure channel; all equalities follow by the definition and the chain rule of mutual information. 

Letting $n\to \infty$, and thus $\epsilon_n\to 0$, establishes the desired converse.


\begin{thebibliography}{10}
\bibitem{maddahali_niesen_2014-1} M. A. Maddah-Ali, U. Niesen, ``Fundamental limits of caching," in \emph{IEEE Trans. on Inf. Theory}, vol.~60, no.~5, pp.~2856--2867, May~2014. 
\bibitem{maddahali_niesen_2014-2} M. A. Maddah-Ali, U. Niesen, ``Decentralized coded caching attains order-optimal memory-rate tradeoff," IEEE/ACM Trans. on Networking, vol.~PP, no.~1, pp.~1, 2014.
\bibitem{niesen_maddahali_2014}  U. Niesen and M. A. Maddah-Ali, ``Coded caching with nonuniform
demands," in \emph{Proc. IEEE INFOCOM Workshop}, 2014.
\bibitem{wang_xian_liu_2015} S. Wang, X. Tian and H. Liu, ``Exploiting the unexploited of coded caching for wireless content distribution," in \emph{Proc. IEEE ICNC, 2015.}
\bibitem{pedarsani_maddahali_niesen_2015} R. Pedarsani, M. A. Maddah-Ali and U. Niesen, ``Online coded caching," \emph{IEEE/ACM Trans. on Networking}, vol.~PP, no.1, pp.~1, 2015.
\bibitem{karamchandani_niesen_maddah_ali_diggavi_2014} N. Karamchandani, U. Niesen, M. A. Maddah-Ali, S. Diggavi, ``Hierar-
chical coded caching," \emph{submitted to IEEE Trans. on Inf. Theory}. Online: \texttt{http://arxiv.org/pdf/1403.7007}.
\bibitem{hachem_karamchandani_diggavi_2014} J. Hachem, N. Karamchandani, and S. Diggavi, ``Content caching and delivery over heterogeneous wireless networks."  Online: \texttt{http://arxiv.org/pdf/1404.6560.}
\bibitem{ghasemi_ramamoorthy} H. Ghasemi and A. Ramamoorthy, ``Improved lower bounds for coded
caching," Online: \texttt{http://arxiv.org/pdf/1501.06003.}

\bibitem{wang_lim_gastpar_2015} C.-Y.~Wang, S.~H.~Lim, and M.~Gastpar, ``Information-theoretic caching:
sequential coding for computing," \emph{submitted to the IEEE Trans. on Inf. Theory.} Online: \texttt{http://arxiv.org/abs/1504.00553v1}.
\bibitem{ji_tulino_llorca_caire_2015} M.~Ji, A.~M.~Tulino, J.~Llorca, and G.~Caire, ``Order-optimal rate of caching and coded multicasting with random demands," \emph{submitted to the IEEE Trans. on Inf. Theory.} Online: \texttt{http://arxiv.org/abs1502.03124}. \bibitem{huang_wang_ding_yang_zhang_2015} W.~Huang, S.~Wang, L.~Ding, F.~Yang, and W.~Zhang, ``The performance analysis of coded cache in wireless fading channel," submitted to \emph{GC2015}. Online: \texttt{http://arxiv.org/abs/1504.01452}. 

\bibitem{Urbanke-1999-A}
R.~Urbanke and A.~Wyner
``Packetizing for the erasure broadcast channel with an internet application,'' 1999. Online:\\ \texttt{http://lthcwww.epfl.ch/\~{}ruediger/papers/inft.ps}. 
\bibitem{Tuncel-Apr-2006-A} E.~Tuncel, ``Slepian-{W}olf coding over broadcast channels,'' \emph{IEEE
  Trans. on Inf. Theory}, vol.~52, no.~4, pp. 1469--1482, 2006.
\end{thebibliography}
\end{document}